\documentclass{aastex62}

\def\kms{\ifmmode{~{\rm km~s^{-1}}}\else{~km s$^{-1}$}\fi}
\def\cm3{\ifmmode{~{\rm cm^{-3}}}\else{~cm$^{-3}$}\fi}
\def\Ms{\ifmmode{~{\rm M_\odot}}\else{M$_\odot$}\fi}
\def\cm3{cm$^{-3}$}

\def\ltsima{$\; \buildrel < \over \sim \;$}
\def\simlt{\lower.5ex\hbox{\ltsima}} 
\def\gtsima{$\; \buildrel > \over \sim \;$}
\def\simgt{\lower.5ex\hbox{\gtsima}} 
\def\arcsec{\hbox{$^{\prime\prime}$}}
\def\deg{\hbox{$^\circ$}}

\def\Chandra{\textit{Chandra}}
\def\PKS{PKS\,1353--341}
\def\ROSAT{\textit{ROSAT}}
\def\Planck{\textit{Planck}}

\usepackage{color}

\accepted{ApJ, March 13, 2019}

\shorttitle{Extended radio and X-ray emissions in PKS\,1353--341}
\shortauthors{Cheung, Giacintucci, \& Clarke}

\begin{document}

\title{Extended Radio Structures and a Compact X-ray Cool-Core in the Cluster Source PKS\,1353--341}

\correspondingauthor{Teddy Cheung}
\email{Teddy.Cheung@nrl.navy.mil}

\author[0000-0002-4377-0174]{C. C. Cheung}
\affiliation{Naval Research Laboratory, Space Science Division, Code 7650, Washington, DC 20375, USA}

\author[0000-0002-1634-9886]{Simona Giacintucci}
\affiliation{Naval Research Laboratory, Code 7213, Washington, DC 20375, USA}

\author[0000-0001-6812-7938]{T. E. Clarke}
\affiliation{Naval Research Laboratory, Code 7213, Washington, DC 20375, USA}

\begin{abstract}

We present a radio and X-ray study of \PKS, the brightest cluster galaxy radio source at the center of a recent \Chandra-discovered X-ray cluster.
Our multi-frequency VLA images reveal an edge-brightened (FR-II), double-lobed structure with total $\sim$50\,kpc extent and 1.5\,GHz power of $1.2\times10^{25}$\,W\,Hz$^{-1}$, separated from the bright, arcsecond-scale core.
We reanalyzed the \Chandra\ data and found the X-ray emitting AGN is offset by $\sim$9\,kpc from a compact X-ray cool-core with temperature, $kT=3.1\pm0.5$\,keV, and a radius of $\sim$22\,kpc, surrounded by a hotter $kT=6.3\pm0.7$\,keV gas out to $\sim$50\,kpc. 
The offset suggests sloshing inside the cool-core induced by a minor merger or a past outburst of the AGN that produced the large-scale radio lobes. 
The comparable spatial scales of the lobes with the interface between the different temperature X-ray plasma indicate the lobes are actively heating the outer layers of what is now a remnant compact cool-core.
Our dual-frequency VLBA images reveal substructure in the central radio source, consisting of a radio core with double-sided pc-scale jets pointing toward the kpc-scale structures. 
The northern jet is detected only at 8.4\,GHz, indicating its emission is behind an absorbing torus or disk.
We also measured faster apparent motions in the southern jet up to $1.9\pm1.1c$ than in the northern jet ($0.8\pm0.5c$).
While the VLBA observations indicate the southern jet is aligned slightly closer to our line of sight, the asymmetries are overall modest and imply minimal projection effects in the large-scale radio structures.

\end{abstract}


\section{Introduction} 
\label{sec:intro}

The radio source \PKS\ is long known to be hosted by a large and luminous galaxy at redshift, $z=0.223$ \citep{whi88,ver00}. 
Because it is a bright \citep[$>$0.5\,Jy;][]{dri97} cm-wavelength flat-spectrum radio source, and its \ROSAT-detected X-ray emission \citep{sie98} is attributable to the central active galactic nucleus (AGN), \PKS\ has been considered a blazar candidate in all-sky catalogs \citep{hea08,mas09}.

Recently, \citet{som18} found through \Chandra\ imaging, that the bright X-ray emitting AGN masqueraded the presence of a massive extended X-ray cool-core galaxy cluster.
The discovery was the outcome of their targeted search for over-densities of red galaxies around such \ROSAT-detected AGN \citep[see also,][]{gre17,yan18}, and confirmed the underlying cluster indicated by the spatially coincident \Planck\ Sunyaev-Zeldovich source PSZ2 G317.79+26.63 \citep[offset of 1.7$'$ from the radio position, thus within the $95\%$ confidence error of 2.4$'$;][]{pla16}. 
Thus \PKS\ can now be classified as the brightest cluster galaxy (BCG) hosting a radio- and X-ray bright AGN at the center of a luminous X-ray cluster.

Interestingly, the \Chandra\ X-ray image of the cluster indicated a weak ($2\sigma$) signature of a pair of cavities at 8.5 kpc (=$2.4\arcsec$) from the BCG \citep{som18}\footnote{Following \citet{som18}, we adopt $H_{\rm 0}=70~$km~s$^{-1}$~Mpc$^{-1}$ and ($\Omega_{\rm M}, \Omega_{\rm \Lambda}$) = (0.3, 0.7), giving a distance of 1.1 Gpc, and angular scale of 3.6 kpc arcsec$^{-1}$ at the source redshift.}. 
Because we found no published radio maps of \PKS\ to compare to the X-ray data, this prompted us to investigate its radio properties using available archival multi-frequency data obtained over many years with the NRAO\footnote{The National Radio Astronomy Observatory is a facility of the National Science Foundation operated under cooperative agreement by Associated Universities, Inc. The VLA is now named the Karl G.~Jansky VLA.} Very Large Array (VLA) and Very Long Baseline Array (VLBA).

The outcome of our study is summarized in the abstract. 
In what follows, \S~2 contains the details of the VLA and VLBA data and our reanalysis of the \Chandra\ data.
In \S~3, the X-ray results are described and compared with the radio structures, while considering constraints on the system from the radio data on all spatial scales.
A discussion and summary is given in \S~4.

All coordinates are given in J2000.0 equinox.
The spectral index, $\alpha$, is defined as $S_{\nu} \propto \nu^{-\alpha}$.
Uncertainties are quoted at the $68\%$ confidence level.

\begin{table*}
  \begin{center}
\caption{Summary of the archival VLA datasets.}
\begin{tabular}{lccccc}
\hline
\hline
Frequency (GHz) & Program$^{\rm a}$ & Dates & Array$^{\rm b}$ & Exposure (s) & $S_{\nu}$ (mJy)$^{\rm c}$ \\
\hline
1.490 & {\it AP142} &1987 Sep 30 &A  &280  &602.4 \\
1.425 & AK509       &2002 Apr 08 &A  &250  &567.3 \\
\hline
4.860 & AT045       &1983 Dec 19 &B/A &1280 &572.6 \\
4.860 & AW136       &1985 Jun 23 &B/C &170  &595.7 \\
4.860 & AD224       &1988 Oct 28 &A  &460  &567.0 \\
4.860 & AK509       &2002 Apr 08 &A  &250  &611.9 \\
\hline
8.460 & AK509       &2002 Apr 06, 08, 09, 10, 21; May 07 & A & 5220 & 703.7 \\
8.460 & AK509       &2002 May 21, 24, 27   &A/B &2335 & $''$ \\
8.460 & AK509       &2003 Feb 09 &D  &370  &674.0 \\
8.460 & {\it AK583} &2005 Jun 08 &B  &405  &663.8 \\
\hline
\end{tabular}
\end{center}
\smallskip
$^{\rm a}$ The two program datasets written in italics were manually calibrated using AIPS. The AP142 dataset used a 21 (out of 27) antenna subset of the array.\\
$^{\rm b}$ Two letters indicate observations taken while antennas were moved between array-configurations.\\
$^{\rm c}$ Measured core flux densities in each dataset.\\
\label{table-1}
\end{table*}

\section{Observations and Data Analysis}

\subsection{VLA Radio}

The archival VLA data for \PKS\ were obtained from 1983 to 2005 while the source was observed as a calibrator in a variety of array configurations (Table~\ref{table-1}). These datasets utilized two 50 MHz wide intermediate frequencies centered at 1.5, 4.9, and 8.5 GHz. Pipeline processing was available for a majority of the cases from the NRAO VLA Archive Survey (NVAS)\footnote{\url{http://archive.nrao.edu/nvas/}} and these calibrated ($u,v$) files were used. Two datasets not in the NVAS (as indicated in Table~\ref{table-1}) were downloaded from the NRAO archive and basic phase and gain calibration was performed using standard procedures in {\tt AIPS} \citep{bri94}. 

After accounting for variability in the core (below) using {\tt UVSUB} in {\tt AIPS}, the ($u,v$) data were combined with {\tt DBCON} to improve ($u,v$) coverage and sampling over a range of spatial scales. The total integration times of the combined 1.5, 4.9, and 8.5 GHz datasets were 9, 36, and 139 minutes, respectively. The data were self-calibrated and imaged using the Caltech {\tt DIFMAP} package \citep{she94} to produce the final images. Because of the southern declination of the source, the resultant beam-shapes of the images were highly elliptical (better east-west resolution). For ease of comparison with the X-ray images, the VLA data were reconvolved with circular beams with different sizes selected to display the structures visible in the {\tt CLEAN} components.

To gauge variability in the core between the different datasets, flux densities were measured in the ($u,v$) plane using {\tt DIFMAP}'s {\tt modelfit} program after an initial round of phase self-calibration was applied assuming an unresolved point source. Comparing measurements at 1.5 GHz and 4.9 GHz that used the same (A-) array revealed modest, but significant variability of $6-8\%$ in the unresolved core flux densities between 1987/1988 and 2002. Because of the identical spatial scales sampled in these data, the variability was likely intrinsic to the source. In the nine 8.5 GHz datasets obtained in April and May 2002 (in A- and A/B-arrays), maximum excursions of only $-1.4\%$/$+2.4\%$ from the average of 703.7 mJy were found, consistent with the $\sim 1-2\%$ systematic uncertainties expected in the absolute flux density scale \citep{per03}; these data were thus combined and treated as a single dataset with $\sim$2-hr total integration. A comparison of the data taken in different arrays indicated $2-5\%$ differences in the 4.9 and 8.4 GHz core flux densities over year-timescales, with no systematic trend of increasing flux with decreased spatial resolution expected if the variability is due to unresolved intermediate-scale structures. Thus for the purpose of combining the different datasets, we accounted for variability in the core to bring its flux (arbitrarily) to the scale of the April 2002 data at 1.5 and 8.5 GHz, and the 4.9 GHz data from December 1983.

\begin{figure}[t]
\begin{center}
\includegraphics[scale=0.75,angle=-90]{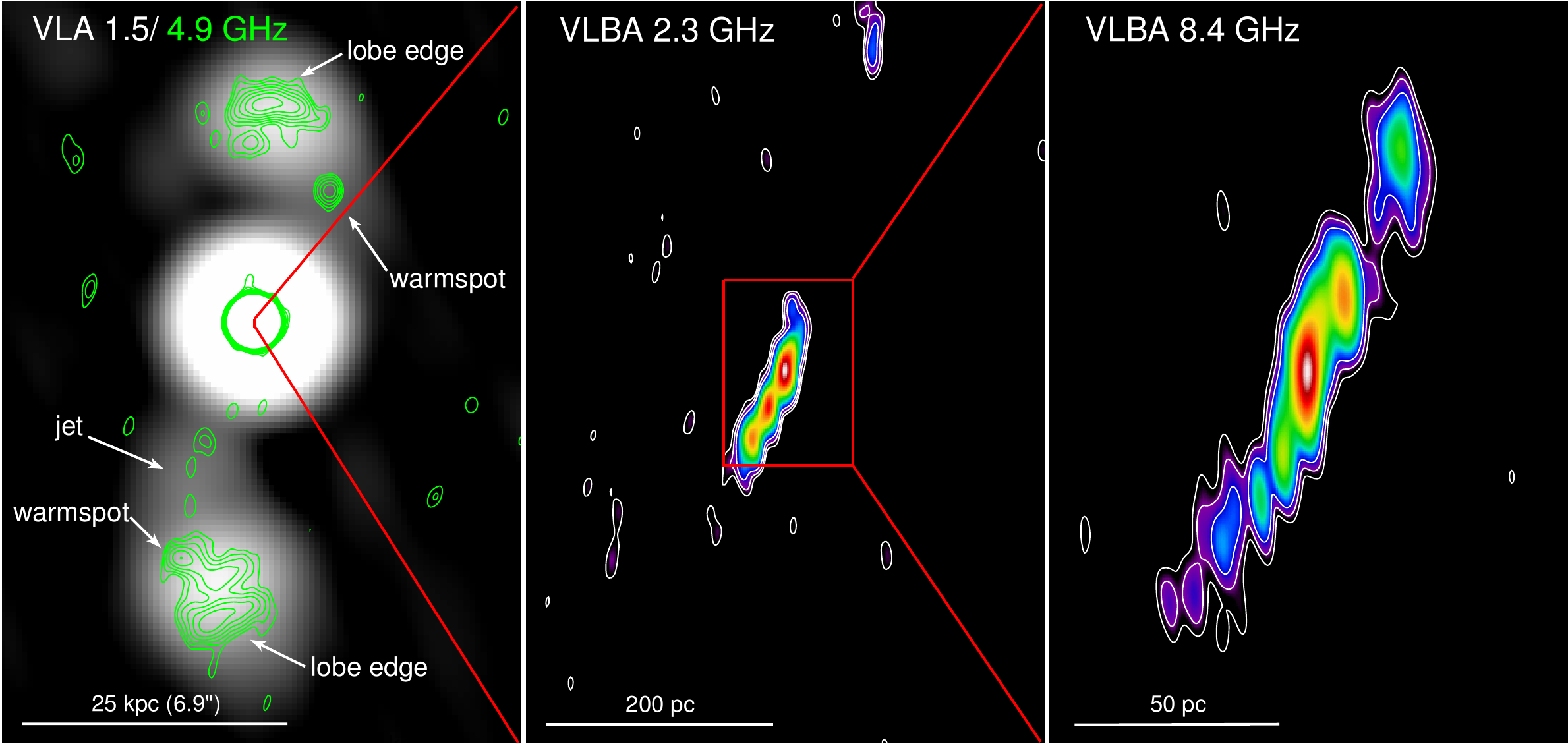}
\caption{Radio images of \PKS\ from kpc-scales to progressively smaller pc-scales from left to right panels. The VLA  images (left panel) at 1.5 GHz (2\arcsec\ beam) is shown in grayscale with green contours from the 4.9 GHz image (0.5\arcsec\ beam; levels from 0.25 to 2 mJy beam$^{-1}$ by factors of $\sqrt{2}$) overlaid. The main features on kpc-scales described in the text are labeled. In the next panels, the two deepest VLBA images from the RFC$^{\ref{rfcfoot}}$ are shown centered on the core, at 2.3 GHz (middle; from January 2015) and 8.4 GHz (right; from April 2012), with colors plus contours at levels 1, 2, and 4 mJy beam$^{-1}$. The beamsizes are 7.1 mas $\times$ 2.8 mas ($PA = -1\deg$) and 3.9 mas $\times$ 1.0 mas ($PA = 6\deg$), respectively.}
\label{radiomontage}
\end{center}
\end{figure}

The main extended arcsec-scale features in the VLA maps of \PKS\ are seen in Figure~\ref{radiomontage} (left) and their measured parameters summarized in Table~\ref{table-2}. Outside of the arcsec-scale radio core, the extended emission is dominated by a pair of edge-brightened lobes (oriented roughly north-south) with warmspots recessed from both lobe-edges; a faint jet connects the core to the southern warmspot. 
For the compact features, we used {\tt DIFMAP}'s {\tt modelfit} program to fit Gaussian components to the data in the ($u,v$) plane. Because the south jet and both extended lobes are resolved, with substantial substructure in the 4.9 and 8.5 GHz maps (see also, \S~\ref{radioXray}), their measurements were made in the map plane. For the features seen at all three frequencies, we found uncertainties in the spectral index of $\sim$0.1.  All other features detected in only the higher-resolution 4.9 and 8.5 GHz images, we measured larger uncertainties in the index of $\sim$0.3.

\begin{table*}
  \begin{center}
\caption{Properties of the extended arcsec-scale radio features.}
\begin{tabular}{lcccccc}
\hline
\hline
Feature$^{\rm a}$ & $d$$^{\rm b}$ & $PA$$^{\rm b}$ & $S_{\nu}$ (1.5 GHz) &$S_{\nu}$ (4.9 GHz) & $S_{\nu}$ (8.5 GHz)  & $\alpha$ \\
              		   & ($\arcsec$) & ($\deg$)  & (mJy) & (mJy) & (mJy) \\
\hline
North lobe edge & 5.7 & --5  & --     &  $8.6 \pm 0.9$    & $4.1 \pm 0.4$ & $1.3 \pm 0.3$ \\
North lobe          & 5.0 & --5  & $34.7 \pm 3.5$ & $12.3 \pm 1.2$  & $6.3 \pm 0.6$ & $1.0 \pm 0.1$ \\ 
North warmspot & 3.9 & --30 & --     & $1.3 \pm 0.1$   & $0.8 \pm 0.1$  & $0.9 \pm 0.3$  \\
\hline
South jet             & 3.3-5.1 	& 158-161 & $4.6 \pm 0.7$ 	& $1.6 \pm 0.2$   & $0.9 \pm 0.1$ & $0.9 \pm 0.1$ \\  
South warmspot	& 6.4 	& 163 	& --  		& $1.6 \pm 0.2$   & $1.0 \pm 0.1$ & $0.9 \pm 0.3$   \\
South lobe  	& 7.3 	& 173 	& $50.9 \pm 5.1$ & $16.7 \pm 1.7$ & $8.8 \pm 0.9$ & $1.0 \pm 0.1$ \\ 
South lobe edge &  7.7 	& 173 	& -- 		& $8.6 \pm 0.9$   & $5.2 \pm 0.5$ & $0.9 \pm 0.3$   \\
\hline
\end{tabular}
\end{center}
\smallskip
$^{\rm a}$ The north and south lobe measurements at all frequencies include the respective lobe edges; the south lobe at 1.5 GHz includes its warmspot.\\
The respective average sizes of the Gaussian fitted dimensions of the north and south lobe edges are ($r$, $a$, $\phi$) = ($1.3\arcsec$, 0.4, $89\deg$) and ($1.6\arcsec$, 0.5, $126\deg$), with the radius of the major axis ($r$), axial ratio ($a$), and position angle of the major axis ($\phi$).\\
$^{\rm b}$ Angular distance ($d$) from the core and position angle ($PA$) defined positive east of north.\\
\label{table-2}
\end{table*}

\begin{table*}
  \begin{center}
\caption{Properties of the milliarcsec-scale radio features.}
\begin{tabular}{lcccccc}
\hline
\hline
Feature$^{\rm a}$	& $d$ & $PA$ & $S_{\nu}$  & $r$ & $a$  & $\phi$ \\
              				& (mas) & ($\deg$)  & (mJy) & (mas) &  & ($\deg$) \\
\hline
\multicolumn{7}{c}{8.7 GHz, 2002 Jan 31} \\
\hline
Core & 0 & 0 & 323.3 & 2.1 & 0.2 & --9 \\ 
N1 & 5.33 & --23.2 & 75.6 & 3.3 & 0.7 & 85 \\ 
\hline
\multicolumn{7}{c}{8.4 GHz, 2012 Apr 05} \\
\hline
Core & 0 & 0 & 248.5 & 3.4 & 0.2 & -9 \\ 
N1 & 5.78 & --24.4 & 120.2 & 3.6 & 0.4 & --4 \\ 
N2 & 16.34 & --22.4 & 45.0 & 3.7 & 0.2 & --7 \\ 
south & 6.16 & 164.4 & 44.1 & 0.9 & 1.0 & 0 \\ 
south & 10.59 & 152.4 & 31.5 & 3.0 & 1.0 & 0 \\ 
south & 18.19 & 151.4 & 10.6 & 0.8 & 1.0 & 0 \\ 
\hline
\multicolumn{7}{c}{8.7 GHz, 2015 Jan 23} \\
\hline
Core & 0 & 0 & 145.7 & 1.4 & 0.4 & --16 \\ 
N1 & 6.10 & --23.5 & 69.4 & 2.1 & 0.5 & 2 \\ 
N2 & 16.77 & --21.2 & 12.5 & 1.3 & 1.0 & 0 \\ 
south & 13.21 & 159.9 & 12.5 & 0.6 & 1.0 & 0 \\ 
\hline
\hline
\multicolumn{7}{c}{2.3 GHz, 2002 Jan 31} \\
\hline
Core & 0 & 0 & 152.2 & 2.5 & 1.0 & 0 \\ 
S1 & 9.83 & 153.7 & 97.7 & 4.1 & 0.2 & --19 \\ 
S2 & 20.00 & 153.8 & 101.1 & 2.8 & 1.0 & 0 \\ 
\hline
\multicolumn{7}{c}{2.3 GHz, 2015 Jan 23} \\
\hline
Core & 0 & 0 & 203.4 & 3.0 & 1.0 & 0 \\ 
S1 & 10.96 & 155.6 & 106.8 & 3.1 & 0.3 & --35 \\ 
S2 & 21.63 & 154.0 & 89.6 & 3.3 & 1.0 & 0 \\ 
\hline
\end{tabular}
\end{center}
\smallskip
$^{\rm a}$ Features in the north (N1, N2) and south (S1, S2) jet are given labels when they are readily identifiable between epochs; otherwise, the jet direction is noted. The remaining columns are as noted in Table~\ref{table-2}. \\
\label{vlbaknots}
\end{table*}

\begin{figure}[t]
\begin{center}
\includegraphics[scale=0.222,angle=-90]{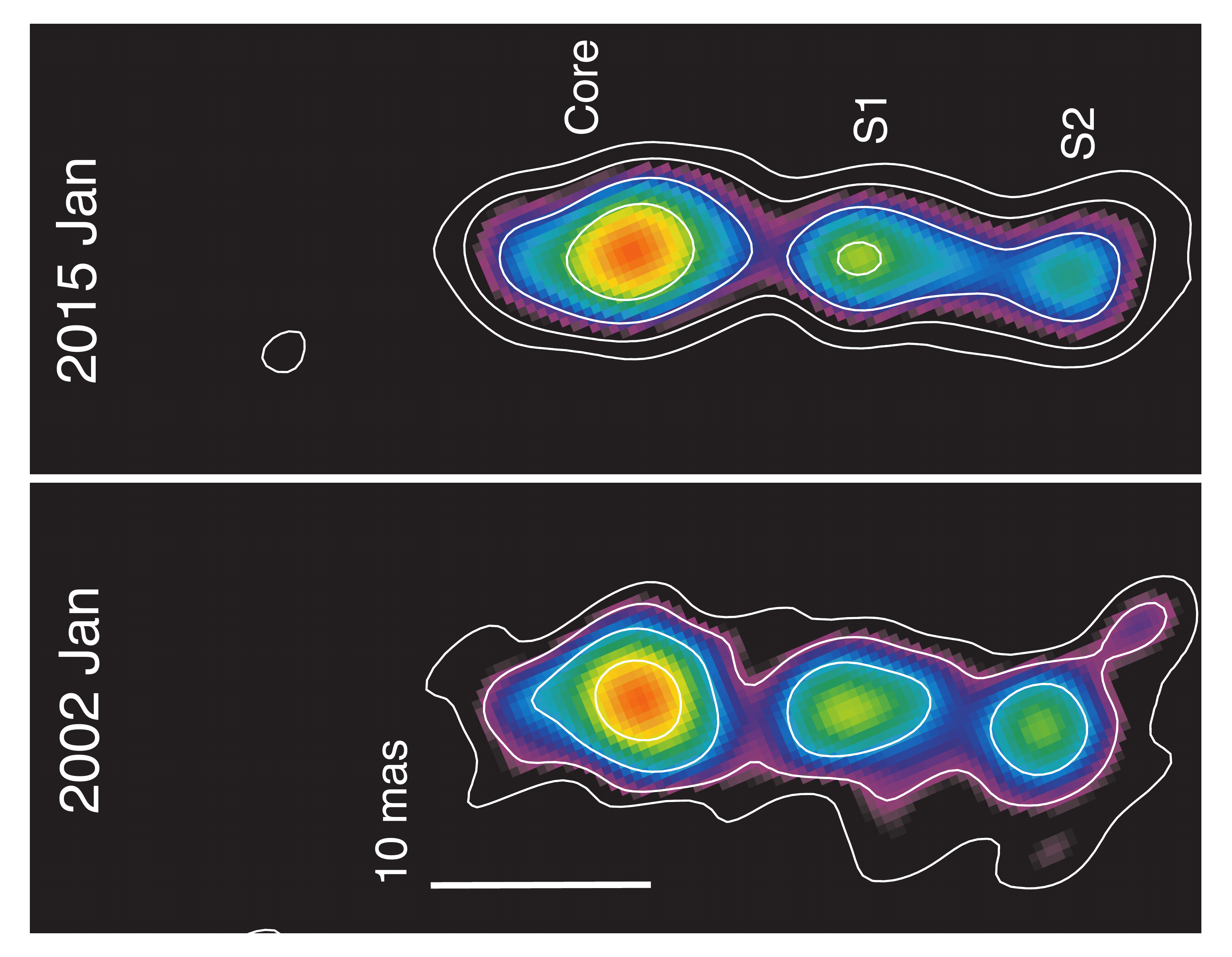}
\includegraphics[scale=0.33,angle=-90]{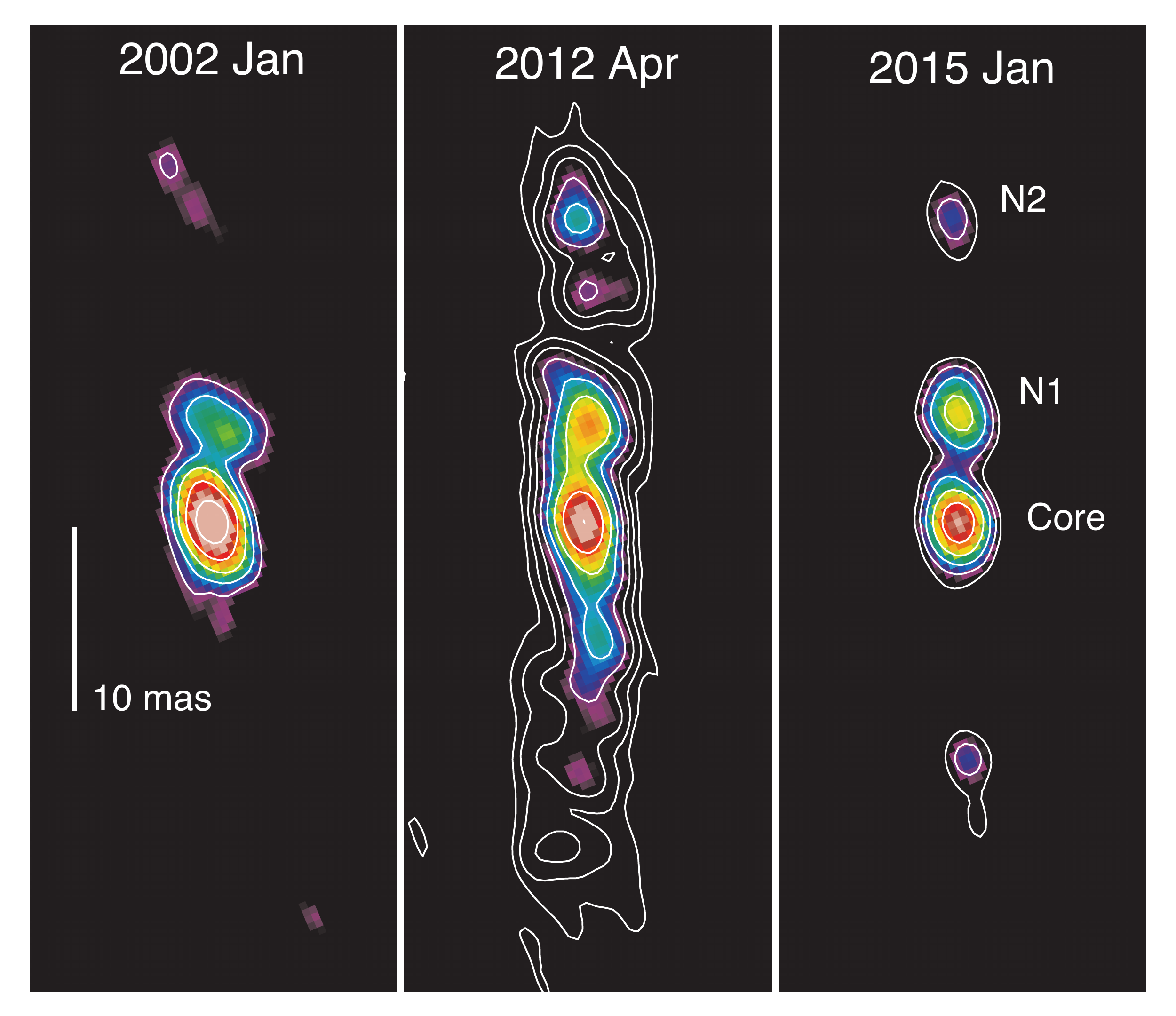}
\caption{VLBA images of \PKS\ rotated by $23\deg$ so the milliarcsec-scale jets are aligned vertically on the figure.
Left two panels are the 2.3 GHz images (4.5 mas beam) with contour levels increasing by factors of two from 8 to 64 mJy beam$^{-1}$.
Rght three panels are the 8.4/8.7 GHz images (2 mas beam) with contours increasing by factors of two up to 128 mJy beam$^{-1}$, beginning at 8, 1, and 4 mJy beam$^{-1}$ (from left to right).
Identifiable features across epochs in the southern (S1, S2) and northern (N1, N2) jets are labeled.}
\label{vlbamontage}
\end{center}
\end{figure}

\subsection{VLBA Radio}

\PKS\ has been observed with the VLBA simultaneously at 2.3 and 8.7 GHz (dual-frequency mode) as part of the VLBA Calibrator Survey's first \citep[VCS-I;][]{fom03} and second \citep[VCS-II;][]{gor16} phases on 2002 January 31 and 2015 January 23, respectively. An additional third epoch imaging at 8.4 GHz was obtained on 2012 April 05 as part of a VLBA imaging survey of nearby radio-bright 2MASS galaxies \citep{con13}. The observations each consisted of two short snapshot scans with total exposures of $\sim 2-5$ minutes. The calibrated VLBA ($u,v$) data and resultant maps from these programs are made publicly available in the Radio Fundamental Catalog \citep[RFC;][]{pet19}\footnote{The  2002, 2012, and 2015 VLBA data used in this work were contributed to the RFC website (\url{http://astrogeo.org/cgi-bin/imdb_get_source.csh?source=J1356-3421}) by Y.~Y.~Kovalev, L.~Petrov, and A.~Pushkarev, respectively. \label{rfcfoot}} and we downloaded them for further analysis.

The low declination of the target resulted in elliptical beam-shapes in the VLBA images (e.g., Figure~\ref{radiomontage}; right two panels) thus to facilitate a comparison, we reimaged all the data and reconvolved the images to a common circular beamsize at each frequency. As seen in the resultant maps (Figure~\ref{vlbamontage}), the data were of varying sensitivity due to VLBA hardware improvements at more recent times.
As expected \citep{gor16}, the 2015 VCS data (200\,s integration) were $\sim$3--5$\times$ more sensitive than in the first epoch in 2002 (300\,s integration) due to 4--12$\times$ greater bandwidths and higher recording rates. The 2012 data (135\,s integration) was observed over the widest frequency range (7.9--8.9\,GHz, centered at 8.4 GHz) resulting in better ($u,v$) coverage at short spacings, thus stands out as showing the best sensitivity to the extended structures.

We used {\tt DIFMAP}'s {\tt modelfit} to fit Gaussian components to the ($u,v$) data to give measured parameters of the different features (Table~\ref{vlbaknots}). Both 2.3 GHz images show a bright core with a single, prominent $\sim$20 milli-arcsecond (mas) long jet pointing toward the southern kpc-scale jet detected in the VLA images. At the same epochs, the 8.7 GHz images show a dominant core plus mainly a faint $\sim$5--6 mas-distant knot to the north. The deeper 8.4 GHz image from 2012 detects both jets, confirming the northern extension up to $\sim$17 mas hinted at in the 8.7 GHz image from 2015, and details the southern jet seen at 2.3 GHz. Estimates of the proper motions of the most distinct features are described in Appendix~A and discussed in \S~\ref{radiocore}. The northern jet is notably absent in the 2.3 GHz images probably because it is behind a surrounding, central disk or torus, and its emission is more-heavily absorbed at lower-frequencies. Taken together, these observations -- the likely absorption-implied orientation of the northern jet, the faster proper motions in the south jet (Appendix~A), and the southern jet detection in the VLA images -- imply the southern jet is aligned slightly closer to our line of sight (\S~\ref{radiocore}).

\subsection{\Chandra\ X-ray \label{chandradata}}

In light of the VLA detections of extended kpc-scale radio emission, we re-examined the \Chandra\ X-ray dataset presented in \citet{som18} that consisted of a 31\,ks observation (ObsID 17214) with ACIS-I in very faint data mode. We reprocessed the level-1 ACIS event file using {\tt CIAO} version 4.8 and the \Chandra\ Calibration Database (CALDB) 4.8.0 following the procedure described in \citet{vik05}. The light curve throughout the observation was inspected \citep[see][]{mar03} and we found no time intervals with significantly elevated background. 

To obtain background-subtracted and exposure-corrected images of the cluster (\S~\ref{Xraycluster} \& \ref{radioXray}), we modeled the detector and sky background using the blank-sky data set from the CALDB appropriate for the date of observation, normalized using the ratio of the observed to blank-sky count rates in the 9.5--12 keV band. We also subtracted the ACIS readout artifact as described in \citet{mar00}.

For regions of interest, we extracted spectra and generated the instrument responses ARF and RMF using the current calibration files for the telescope effective area, CCD quantum efficiency, and ACIS time-dependent low-energy contamination model. The emission from a $3\arcsec$ radius region coinciding with the central X-ray AGN found by \citet{som18} was excluded from our analysis of the extended cluster emission. For the same regions, we extracted the background spectra from the corresponding black-sky data sets, normalized as described above. The spectra of the extended X-ray emission was fitted in {\tt XSPEC} 12.9.0 with an absorbed, single-temperature APEC model in the 0.8--7\,keV energy band, with the metal abundance left free to vary and the column density of HI absorption fixed to the Galactic value of $5.57\times10^{20}$ cm$^{-2}$ \citep{kalb05}.

\begin{figure}[ht]
\begin{center}
\includegraphics[scale=0.47,angle=0]{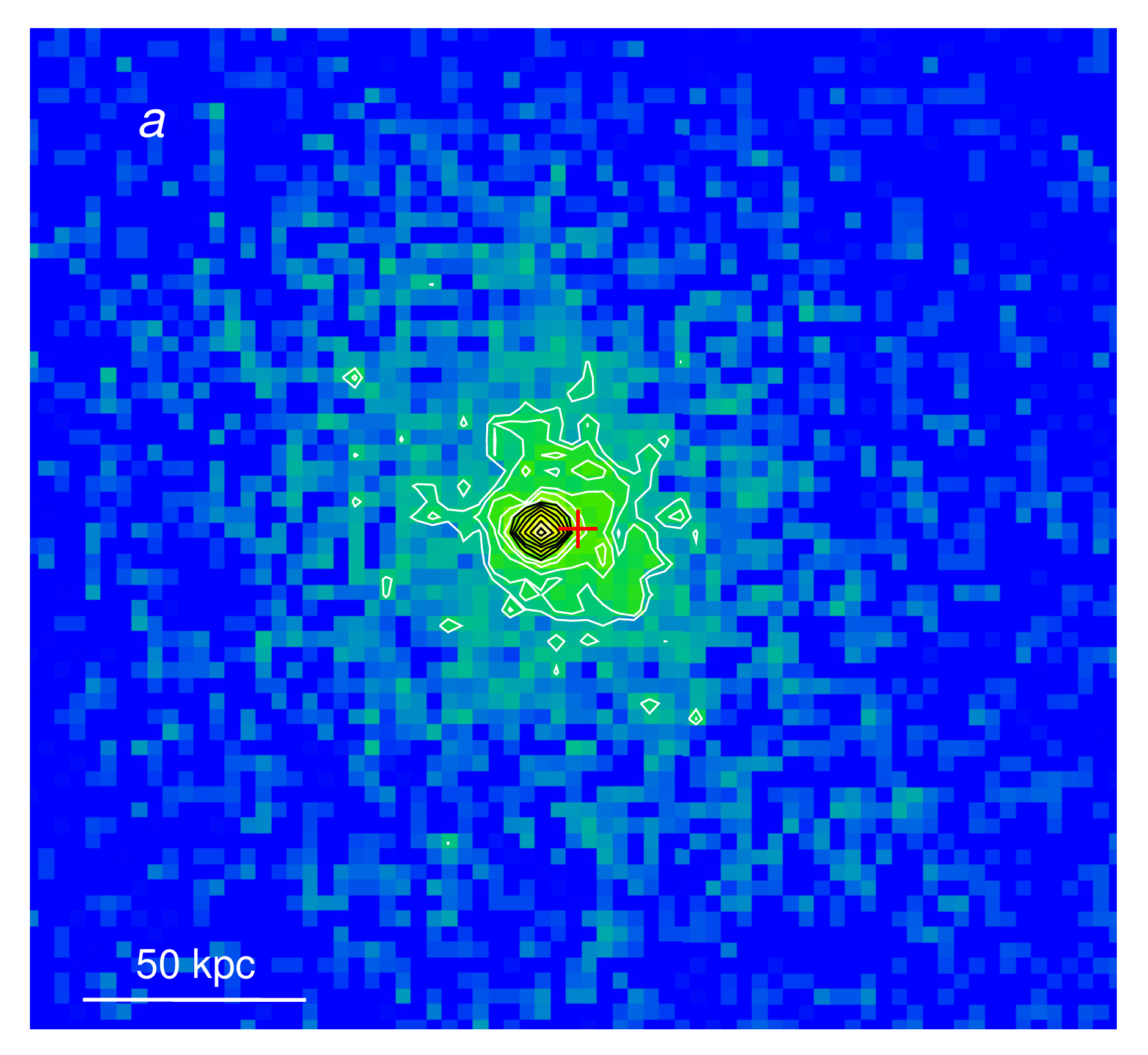}\hspace{-0.15in}
\includegraphics[scale=0.47,angle=0]{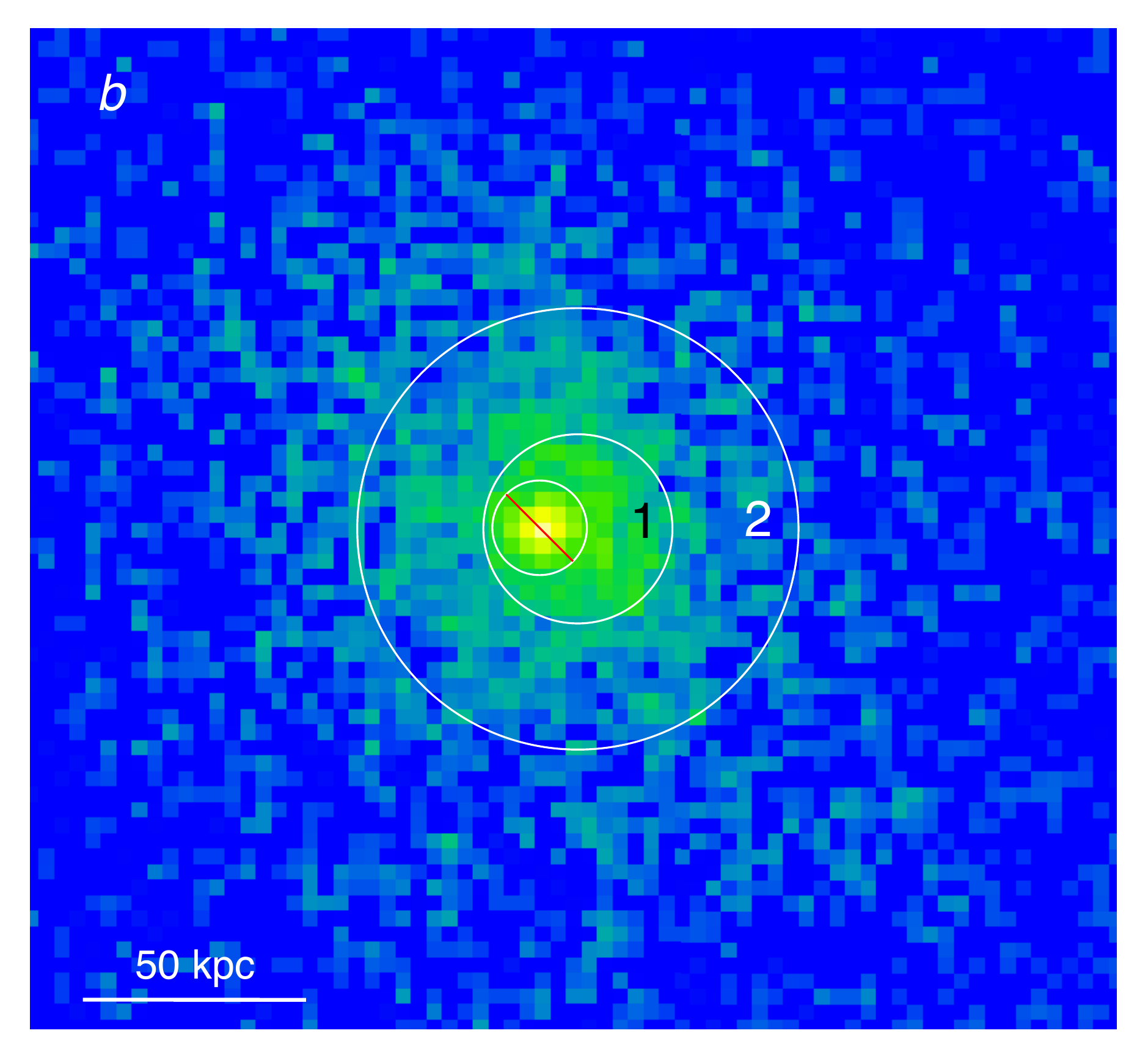}
\includegraphics[scale=0.55,angle=0]{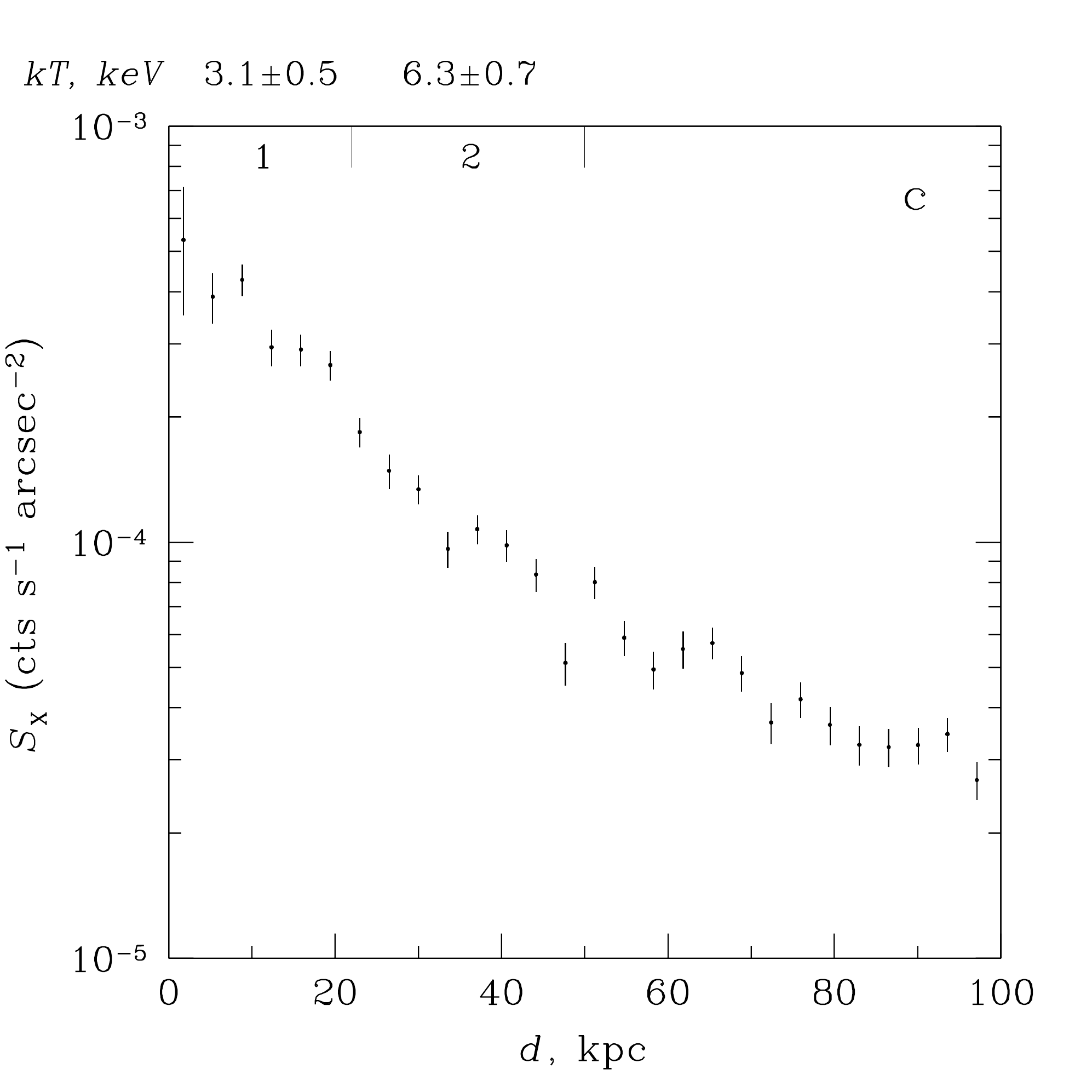}
\caption{{\em Top panels (a, b)}: \Chandra\ X-ray image of \PKS\ in the 0.5--4 keV band, binned to 2 pixels ($\sim 1\arcsec$) and smoothed with a Gaussian of 1 pixel ($\sim$$0.5\arcsec$) width. In panel {\em (a)}, X-ray contours of the central brightest region (from the same \Chandra\ image) are overlaid with levels increasing from the outermost one ($1.8\times10^{-4}$ counts s$^{-1}$ arcsec$^{-2}$) toward the center by factors of $\sqrt{2}$. The black contours correspond to the AGN X-ray point source that is offset by $\sim$$2.5\arcsec$ (9 kpc) from the center (red cross) of the bright, extended core with $r \sim 22$ kpc (white contours). In panel {\em (b)}, white circles indicate the regions used for spectral extraction, with emission from the AGN ($r=3\arcsec$ circle with red line) excluded. Panel {\em (c)} shows the surface brightness profile extracted from the full-resolution 0.5--4 keV image. The fitted temperatures in the corresponding regions marked in panel {\em (b}) are indicated.
}
\label{chandramaps}
\end{center}
\end{figure}

\section{Results}

Here, we revisit the X-ray spatial and spectral analysis of the inner 100-kpc radius of the cluster emission that coincides with our newly found radio structures (\S~\ref{Xraycluster}). 
The X-ray results enable a comparison of the extended radio and X-ray structures (\S~\ref{radioXray}).
We then discuss the VLBA observations of the inner unresolved radio core and pc-scale jet features, including their measured apparent proper motions, in the context of the historical integrated radio spectrum (\S~\ref{radiocore}).

\subsection{Central X-ray Cluster emission \label{Xraycluster}}

A \Chandra\ image in the 0.5--4 keV band is shown in Figure~\ref{chandramaps}. 
We overlay a set of contours (from the same image) to show the finer-structure in the X-ray emission in the central region of the map.
The X-ray point source coinciding with the AGN (black contours) found by \citet{som18} coincides with the radio core in our VLA and VLBA images. 
The white contours highlight the presence of a relatively compact region of radius $\sim$22\,kpc that encloses the brightest extended X-ray emission in the cluster center.
The extended emission is brightest to the west of the X-ray AGN, and we find an offset of $\sim$$2.5\arcsec$ ($\sim$9\,kpc) between its centroid (red cross) and the AGN.

In Figure~\ref{chandramaps}{\em c}, we show a \Chandra\ X-ray radial surface brightness profile out to 100 kpc radius extracted with annular regions centered on the centroid of the extended emission, excluding a $r=3\arcsec$ region around the AGN, using the full-resolution 0.5--4 keV image.
There is a hint of a change in the surface brightness at $r \simlt 22$ kpc (region 1), coincident with the bright central core we identified by-eye in the X-ray image contours. 
Both the profile and image suggest a possible sharp brightness edge delineating an enhanced, inner compact core (at least in parts of the azimuth), however the statistics are insufficient to rule out a smooth boundary. 

We fitted the temperature (\S~\ref{chandradata}) of this compact core delineated by region 1 ($r=22$ kpc) and the surrounding gas within region 2 (22 kpc $< r <$ 50 kpc) indicated in Figure~\ref{chandramaps}{\em b}. 
The compact core is significantly cooler than the outer annulus, with (source rest-frame) temperatures of $kT = 3.1\pm0.5$ keV and $6.3\pm0.7$ keV, respectively (also reported in Figure~\ref{chandramaps}{\em c}). 
The significantly different temperatures derived support the idea that the region within $\sim$22\,kpc is a physically distinct emission component from its surrounding annulus of $\sim$22--50\,kpc.
These results also indicate that the \PKS\ cluster lacks a large-scale cool core, typically found in many relaxed clusters \citep[e.g.][]{hud10}.
For the inner region 1, we found a 0.5--2.0 keV luminosity, $L_{\rm X} = (1.2 \pm 0.1) \times 10^{43}$ erg s$^{-1}$.

The X-ray surface brightness profile of the galaxy cluster studied by \citet[][Figure~4, therein]{som18} was centered on the AGN, and is thus systematically offset by $\sim 2.5\arcsec$ = 9 kpc from our profile that was centered on the compact, extended X-ray core.
Because we found a significant offset between the AGN point source and the center of the compact core, we were able to study the details of the surface brightness profile at radii within 50 kpc, thus enabling the detection of the $kT \sim 3$ keV component within 22 kpc.
According to their profile, the wings of the AGN point spread function has only a small to negligible effect on our results because its surface brightness is at $\sim$$10\times$ smaller level than the diffuse emission at the radius of $r=3\arcsec \sim 11$ kpc we used for its excision and drops rapidly with increasing distance.
Their profile analysis extends out to radius $\simgt$1 Mpc, and at radii from 50 kpc and greater, our profiles should be sufficiently similar to utilize their fitted results there. 
They found the statistics were sufficient to determine the temperature profile from 50--700 kpc that appears consistent with a constant temperature, $kT \sim 8$ keV. 
At smaller radii (in the single bin analyzed from 10--50\,kpc), they found a lower-temperature of $\sim$5.5\,keV, consistent with our measurement in region 2.

\begin{figure}[t]
\begin{center}
\includegraphics[scale=0.5,angle=0]{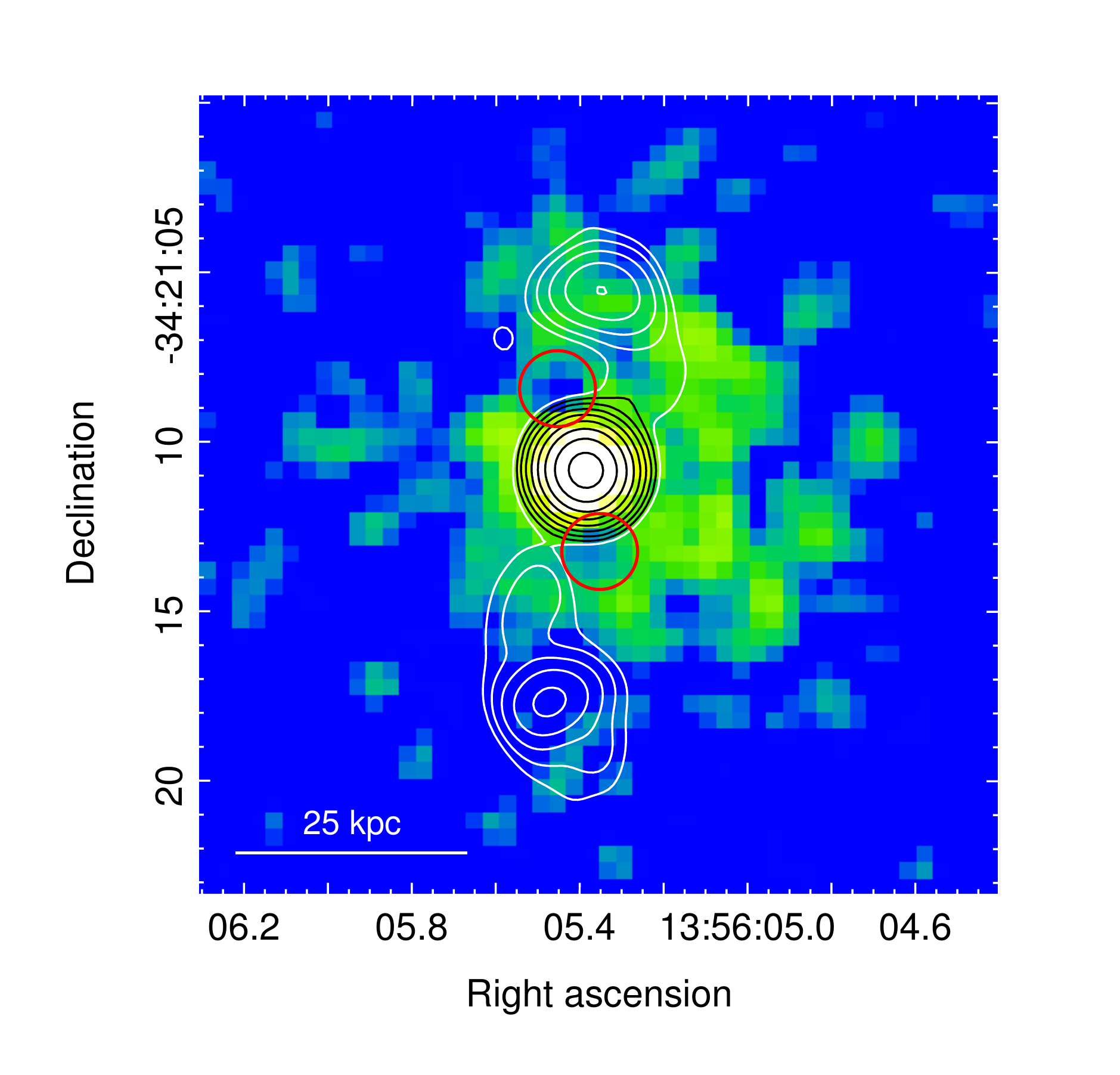}\hspace{-0.5in}
\includegraphics[scale=0.5,angle=0]{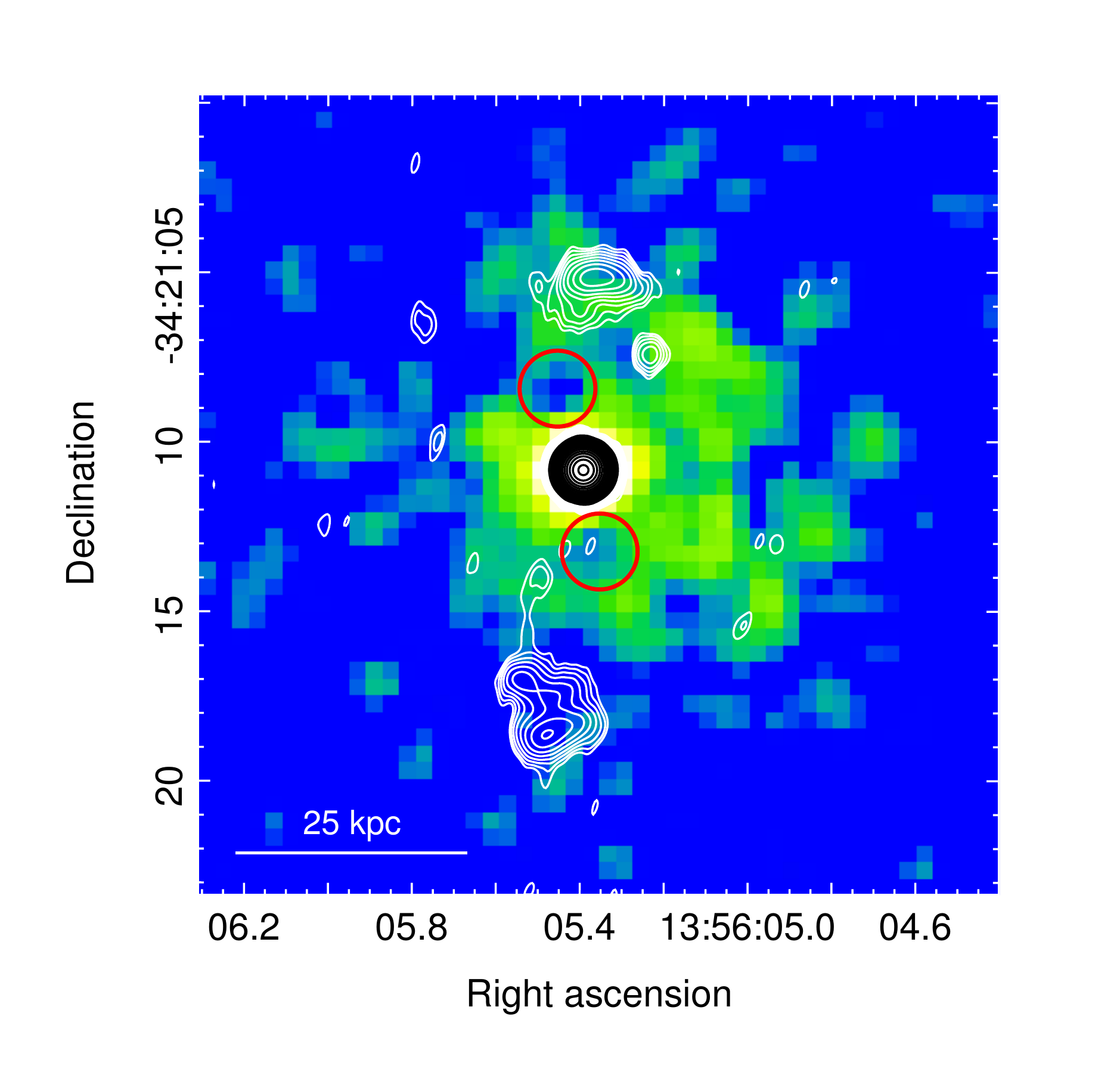}
\\
\vspace{-0.25in}
\includegraphics[scale=0.5,angle=0]{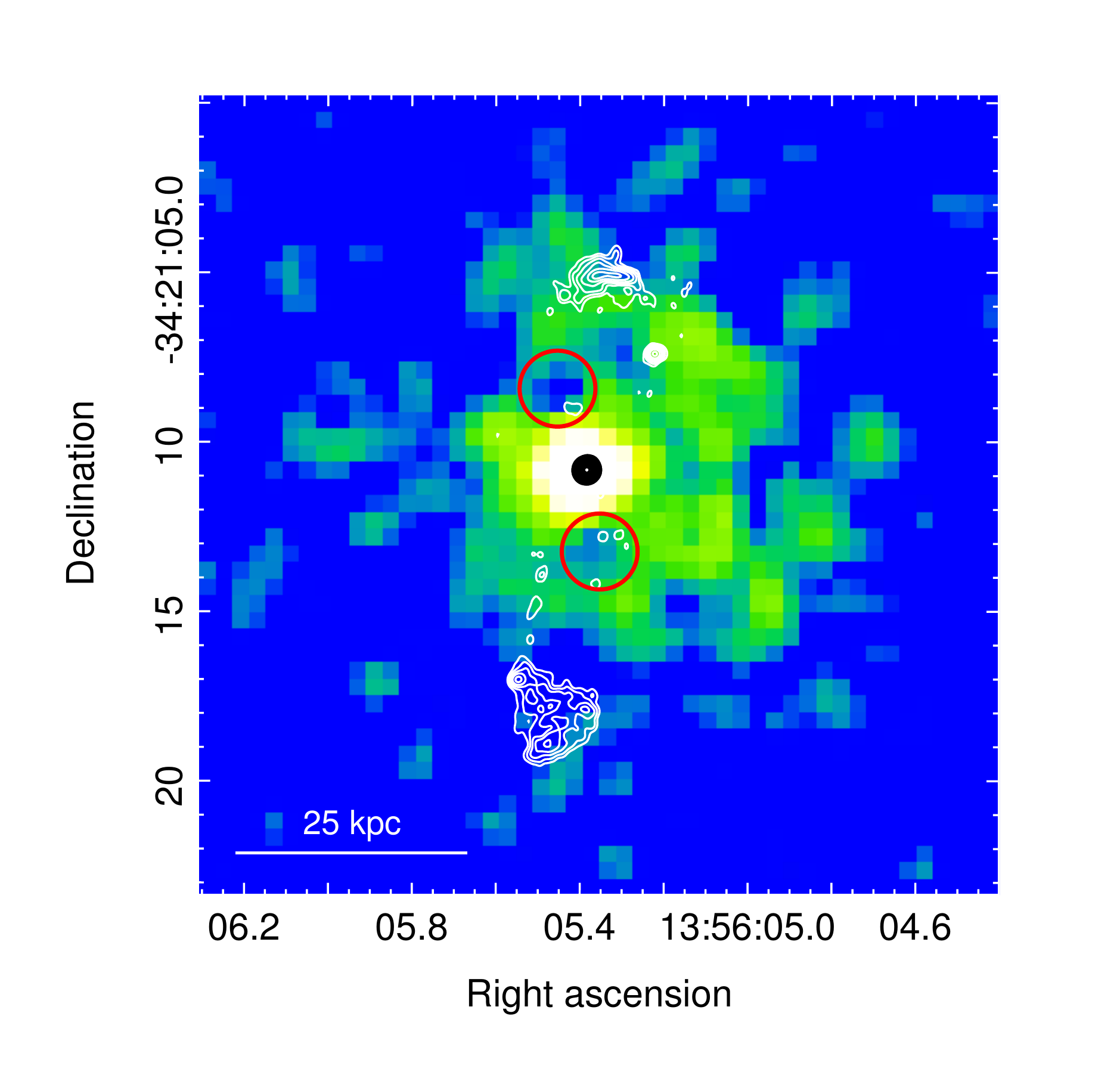}
\caption{\Chandra\ X-ray image of \PKS\ in the 0.3--2 keV band (colors) overlaid with the contours from the VLA 1.5 GHz (1.5\arcsec\ beam; upper left), 4.9 GHz (0.75\arcsec; upper right), and 8.4 GHz (0.3\arcsec; bottom) images. The minimum contour levels are 1.5, 0.25, and 0.1 mJy beam$^{-1}$, respectively, and increase by factors of 2, $\sqrt{2}$ and $\sqrt{2}$. Red circles mark the positions of the weak X-ray cavities found by \citet{som18}.}
\label{radioXrayoverlay}
\end{center}
\end{figure}

\subsection{Kpc-scale Radio and X-ray Emission \label{radioXray}}

A \Chandra\ X-ray image at softer energies (0.3--2 keV band) is presented  in Figure~\ref{radioXrayoverlay} with the VLA radio contours at 1.5, 4.9 and 8.5 GHz overlaid. The main features visible in the X-ray map are the possible weak ($2\sigma$) X-ray decrements 2.4\arcsec\ = 8.5 kpc from the AGN (marked with red circles in Figure~\ref{radioXrayoverlay}) reported by \citet{som18} that prompted our study of this system. 
These suspected cavities are found within the highest surface brightness X-ray emission of the central cluster emission characterized by the $kT \sim 3$ keV temperature.
No radio emission is found inside the cavities in our images.
In fact, the cavities axis ($PA = 15\deg$ and $190\deg$), is misaligned with respect to the radio axis on kpc-scale ($PA = -30\deg$ and $-5\deg$ to the north and $158\deg -173\deg$ to the south) and pc-scales ($PA = -23\deg$ and $154\deg$).
The cavities' proximity to the cluster center, combined with the larger radial distance of the double-radio lobes, makes it unlikely that they are older relic ghost bubbles inflated by a past outburst of the AGN \citep[e.g.,][]{fab12}. 

The pair of recessed, compact warmspots detected in the highest-resolution VLA images (0.3\arcsec) help to define the radio source axis. 
While a faint jet points to the southern warmspot, the northern jet is undetected in our images. 
We call them warmspots \citep[see][]{sau81} to emphasize that they are not high brightness temperature features as seen in classical \citet{fan74} type-II (FR-II) radio doubles, but do seem to mark the active surface at the end of the kpc-scale jets \citep[e.g.,][and references therein]{bri94b}.   
The northern and southern warmspots do not have obvious counterparts in the optical images of \citet{som18}, so they are not obviously field radio sources.
Note in \citet{ver00}, there is an optical source nearest the BCG center that is a galaxy 4.8\arcsec\ to the N/NE at $PA = 19\deg$ with the same redshift, but we find no radio counterpart to the source.
Both lobe edges look edge-brightened and the northern one in particular looks shell-shaped perhaps indicating enhanced interaction with the surrounding medium in this direction.
The change in direction from the warmspots to the lobes gives the radio source an overall `Z'-shaped inverse symmetry that could indicate a precession of the central source, or could be due to interactions with the surrounding medium.
The symmetry extends down to smaller scales (\S~\ref{radiocore}), with modest, but significant differences in position angles of the structures from kpc- to pc-scales of $\sim$$7\deg$ (north) and $\sim$4--9$\deg$ (south). 

The limited statistics of the current \Chandra\ exposure makes it difficult to determine if the X-ray surface brightness distribution in the 2-dimensional map at radii $\simgt 22$ kpc shows any significant features directly associated with the extended radio structures, particularly because the outer region is characterized by a hotter $kT \sim 6$ keV temperature (less soft photons).
Broadly, spatial comparisons between the radio and X-ray emissions are likely minimally affected by projection effects (\S~\ref{radiocore}).
We find the entirety of the southern lobe (angular distance of the inner to outer edges of $\sim$23--28\,kpc) lies just outside of the central $kT \sim 3$ keV emission, and is embedded in the hotter $kT \sim 6$ keV gas (\S~\ref{Xraycluster}).
The north lobe edge at an angular distance of $\sim$21\,kpc coincides more closely with the interface of the between the central $kT \sim 3$\,keV ISM and hotter surrounding gas. 
Otherwise, there is little sign of preferred enhanced activity / interaction of the radio jets and lobes with the surrounding gas.

\begin{figure}[t]
\begin{center}
\includegraphics[scale=0.85,angle=0]{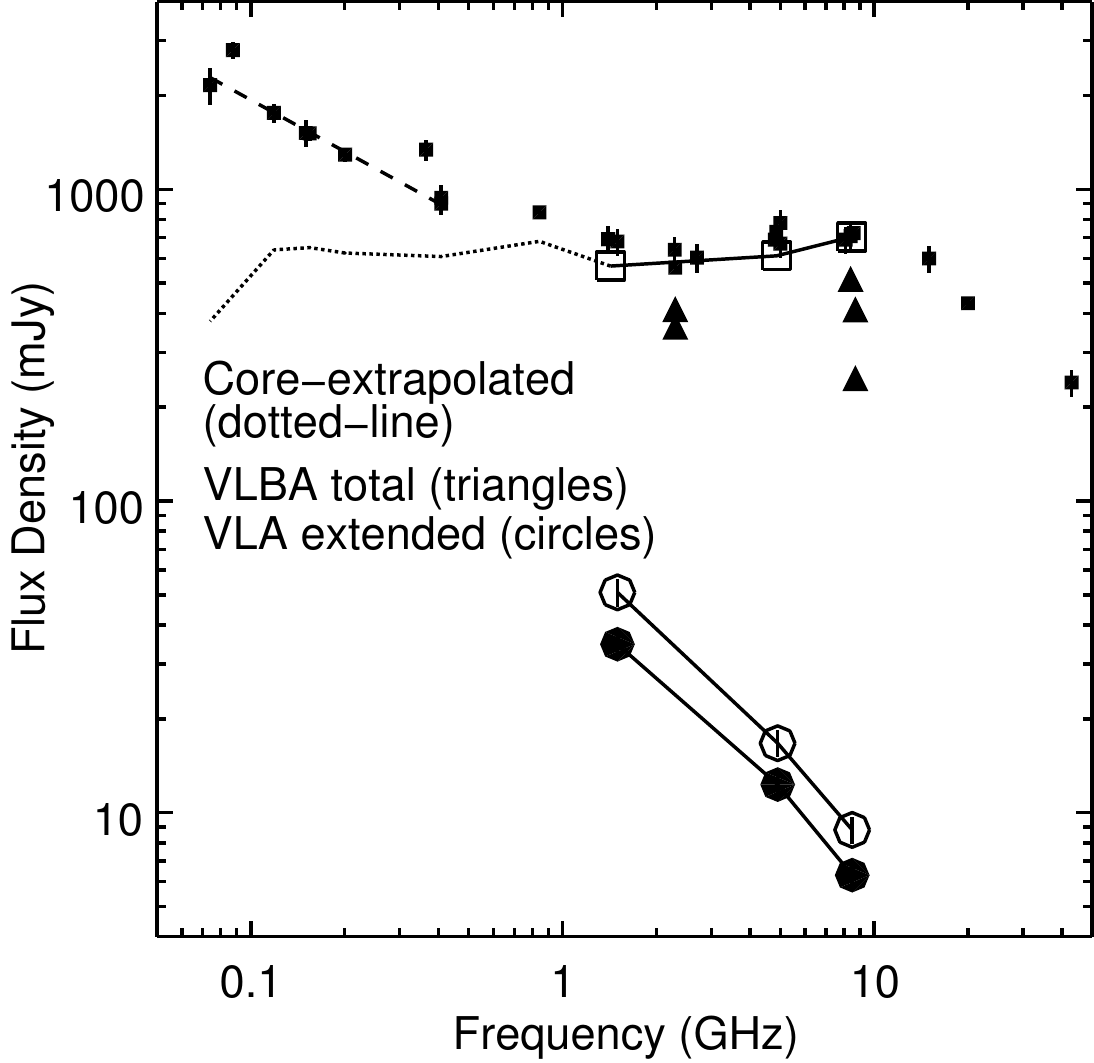}
\caption{Radio spectra of different components of \PKS. 
The integrated measurements are the filled small squares (Appendix~B). The dashed line shows the best-fit slope, $\alpha = 0.55$ between 74 and 408 MHz. Other measurements include the total VLBA fluxes from Table~\ref{vlbaknots} (triangles) and the VLA extended emission (north = filled and south = open circle). The simultaneous VLA data points for the arcsec-scale radio core from April 2002 (Table~\ref{table-1}) are shown as larger open squares (errors smaller than symbol size) with the extrapolated core fluxes at lower frequencies assuming the power-law spectra of the extended VLA features continue unbroken down to 74 MHz (see text).}
\label{radiosed}
\end{center}
\end{figure}

The extended components found in the VLA 1.5 GHz image constitutes $\sim$$15\%$ of the total 1.5 GHz flux and the northern and southern lobes totaled correspond to a radio power of $1.2 \times 10^{25}$ W Hz$^{-1}$.
The radio spectra of the lobes and its fainter features (jet, warmspots) are all consistent with a single power-law spectral index, $\alpha \sim 1$ in the VLA frequency bands (Table~\ref{table-2}).
As a prelude to the next subsection, it is interesting to try to reconcile the relationship between the integrated flux measurements at $<$1\,GHz frequencies seen in the integrated spectrum (\S~\ref{radiocore}) to consider the possible low-frequency extension of the spectra of these extended components.
The low-frequency integrated spectrum from 74 to 200.5 MHz is consistent with a single power-law ($\alpha = 0.55 \pm 0.12$) with the highest-resolution measurement taken at $25\arcsec$ from the GMRT TGSSADR (Appendix~B). 
The spectrum extrapolates to the two 408 MHz integrated measurements, but under-predicts all the higher-frequency ones (Figure~\ref{radiosed}) indicating a separate spectral component at $\simgt 1$ GHz dominated by the VLA core.
Connecting the low-frequency (74--408 MHz) integrated spectrum to the spectrum of the extended VLA components would require a large break of $\Delta\alpha \sim 1.1$, meaning $\alpha$ (0.4--1.5 GHz) $\sim 1.7$ that would be inconsistent with the $\alpha \sim 1$ found for the extended components at 1.5--8.5 GHz, amounting to a very unusual overall saw-tooth spectral shape.
Instead, because we find the $\simgt 1$\,GHz integrated measurements are actually a superposition of many unresolved sub-arcsec scale components (\S~\ref{radiocore}), thus favoring the interpretation that the VLBA-scale components' spectra extend to lower energies. Indeed, extrapolating the 1.5--8.4 GHz spectra of the two lobes plus the southern jet with a single unbroken power-law down to 74 MHz, and subtracting this from the integrated measurements defining the $\alpha = 0.55$ integrated component, we obtain a residual (dotted line in Figure~\ref{radiosed}) that shows a reasonable flat spectrum from 74 MHz to 43 GHz for the central, sub-arcsec component.
Taking the extrapolated total radio lobe flux at 150 MHz of 849 mJy, the total radio power, $L_{\rm 150 MHz} = 1.2 \times 10^{26}$ W Hz$^{-1}$ is on the upper end observed in systems with jet powers derived from X-ray cavities by \citet{kok17}. 
Systems with similar or greater 150 MHz power have jet powers $\sim$(1--70)$\,\times 10^{44}$ erg s$^{-1}$, 
with a value of  $3 \times 10^{44}$ erg s$^{-1}$ implied by the relation found in their ensemble sample.

\subsection{The Central Radio Source \label{radiocore}}

Existing radio studies of \PKS\ were limited to published integrated flux measurements of the source \citep{som18}.
The NVSS 1.4 GHz flux density of 691.8 ($\pm 20.8$) mJy (obtained in May 1996) is consistent with our total VLA  fluxes at the same frequency band (core fluxes in Table~\ref{table-1} plus extended fluxes in Table~\ref{table-2}), and implies a total radio power of $1.0 \times 10^{26}$ W Hz$^{-1}$. 

Our compilation of all the integrated measurements of the source (Appendix~B; Figure~\ref{radiosed}) shows a steep-spectrum component at low-frequencies and the flat-spectrum component at $>$1\,GHz known from previous studies \citep[e.g.,][]{jau82,hea08,mur10}. 
The overall radio spectrum appears to match one of the spectral templates of \citet[][Figure~1, therein]{hog15}, with what initially appears to be a $>$1\,GHz gigahertz-peaked spectrum \citep[GPS;][]{ode98}.
However, our analysis in \S~\ref{radioXray} demonstrated the spectrum of the unresolved arcsec-scale core that dominates the $>$1\,GHz emission likely extends down to the lowest-observed frequencies, indicating an underlying origin from a superposition of different aged emission features. 

Indeed, our VLBA maps measured mas-scale core fluxes at $\sim$2--8 GHz that fall well below ($<$1/2) the arcsec-scale radio fluxes.
Considering also the jet knots, a sum of all the VLBA-detected components (plotted in Figure~\ref{radiosed}) still lie below those of the unresolved arcsec-scale VLA core flux density measurements, indicating there is further intermediate-scale emission not detected in these VLBA data.
Note the RFC database$^{\ref{rfcfoot}}$ results indicate even finer-scale structure in these data with unresolved flux densities of 44--70 mJy at 8.4/8.7 GHz and 55--70 mJy at 2.3 GHz.
While the arcsec-scale core in our VLA data is mildly-variable (up to $8\%$; \S~2), the 2.3/8.7 GHz spectrum of the mas-scale core is strongly variable. The spectrum of the latter remained optically thick over the 13-years, changing from being strongly inverted ($\alpha = -0.6$) to flat  ($\alpha = 0.3$) mostly due to its decline in flux density at 8.7 GHz from $\sim$320 mJy to $\sim$150 mJy. 

The best 8.4 GHz image (Figure~\ref{vlbamontage}) detects both jets out to $\sim$20 mas in both the north and south directions.
While there appears to be an additional, very faint feature at $\sim$90 mas ($PA = -15\deg$) further to the north in the RFC 2.3 GHz map (Figure~\ref{radiomontage}, center panel), the absence of the north jet in the 2.3 GHz images at smaller distances indicate the northern jet is apparently heavily absorbed at the lower frequency 2.3 GHz maps.
At the position of N2, we measured a $3\sigma$ rms limit of $<6$ mJy in the (deeper 2015) 2.3 GHz image, that translates to a constraint on the spectral index of $\alpha \simlt -0.6$, consistent with a heavily-absorbed spectrum. Note the AT20G catalog \citep{mur10} indicates very low polarization for the arcsec-scale core, $1.1\%$ at 8 GHz, and upper limits of $<$$3.1\%$ at 20 GHz and $<$$1.1\%$ at 5 GHz, confirming the presence of dense material in the inner regions.

With the long time-baseline of the observations, we derived proper motions for the features in the VLBA maps that are identifiable between epochs (Appendix~A). 
At 2.3 GHz, the VLBA images show the southern jet extends out to $\sim$20 mas, with the outer knot S2 moving radially outward ($PA = 154\deg$) with apparent velocity, $\beta_{\rm app} = v/c =1.9 \pm 1.1$.
The inner feature, S1, has more elongated structure in the maps with marginal evidence of outward motion, $\beta_{\rm app} \sim 1.3 \pm 1.2$, and no significant change in flux.
The only-measured proper motion in the north jet is for N1, at  $PA = -24\deg$, with $\mu = 0.055 \pm 0.034$ mas yr$^{-1}$ that implies $\beta_{\rm app} = 0.8 \pm 0.5$.

Assuming constant velocities over time, the inferred epochs of zero-separation (i.e., kinematic ages) of N1 and S1 are both $\sim$10$^{2}$ years prior to the first VLBA epoch in 2002, with ranges of 60--250 and 60--980 years, respectively.
Because of its larger proper motion, the more distant S2 knot gives a range of $(1-4)\times10^{2}$ years that is similar to the other features, and also helps to limit the upper uncertainty in the age of S1 (it must be younger than S2).
Taking the respective proper motions of S1 and N1 as the approaching ($\mu_{\rm a}$) and receding ($\mu_{\rm r}$) jets, they provide only a weak constraint on the quantity \citep{ree66,mir99}, $\beta_{\rm p}\, {\rm cos}\,\theta = (\mu_{\rm a}-\mu_{\rm r})/ (\mu_{\rm a}+\mu_{\rm r}) = 0.24 \pm 0.51$, thus unfortunately no useful constraints on the pattern velocity ($\beta_{\rm p}$) and line-of-sight angle.
Overall, the kinematic analysis indicates the pc-scale structures are relatively young, with likely ages of order $\sim$10$^{2}$ years.

Despite their large uncertainties, the modest values of the proper motions imply the pc-scale structures lie close to the plane of the sky, consistent with the roughly symmetric brightnesses of the two jets in the deep 8.4 GHz image.
We can infer from the faster motions in the southern mas-scale jet and the possible absorption of the northern jet that the southern axis is slightly more closely aligned to our line of sight.

\section{Discussion and Summary \label{summary}}

We studied the kpc- to pc-scale radio structures of \PKS\ ($z=0.223$), the BCG at the center of a recently discovered X-ray cool-core cluster by \citet{som18}.
Our VLA images reveal the large-scale radio emission exhibits an edge-brightened (FR-II) radio morphology.
They have a radio power at 1.5 GHz of $1.2 \times 10^{25}$ W Hz$^{-1}$, that is near the dividing line between FR-I and -II \citep{wol07,che09}.
FR-II radio sources at the centers of nearby clusters are relatively rare, being found in $<1\%$ of Abell clusters at $z < 0.25$ \citep{owe92,owe97,sta14,hag15}; see also e.g., Cygnus~A \citep[e.g.,][]{arn84,smi02}.
Taking the FR-II classification together with the optical spectrum of the active galactic nucleus with narrow-emission lines typical of a Seyfert type-2 galaxy \citep{ver00}, makes it best-classified as a narrow-line radio galaxy \citep{vei87}; see also \citet{cur06}. 

On pc-scales, the VLBA-observed emission consists of a two-sided jet aligned toward the kpc-scale structures, and centered on a flat-spectrum radio core.
The detection of the northern jet at only the higher frequency is analogous to other well-studied radio galaxies where the faintness of the counter-jet at the lower-frequencies is associated with free-free absorption due to the presence of a torus or disk \citep[e.g.,][]{kri98,wal00}. 
Indeed, such a torus or disk with radius of 10's pc or more was indicated in the detected $HI$ absorption \citep{ver00}. 
The inferred absorption of the northern jet makes it likely to be the counter-jet (aligned away from our line of sight), consistent with its smaller (sub-luminal) proper motions, $\beta_{\rm app} = 0.8 \pm 0.5$, than observed in the southern jet (maximum of $\beta_{\rm app} =1.9 \pm 1.1$).

Our reanalysis of the spatial and spectral properties of the central 100\,kpc of the X-ray cluster in the \Chandra\ data suggest that the presence of a central bright X-ray structure with a radius, $r \sim 22$ kpc, and temperature of $\sim$3 keV.
Despite being contained within the optical extended envelope of the host BCG that is known to be particularly large (extent $\simgt$400 kpc) and luminous \citep{ver00}, with $L_{\rm K}/L_{\odot} = 1.6 \times 10^{12}$ \citep[for $K_{\rm s}$-band magnitude of 13.1;][]{skr06}, the feature is inconsistent with being a galactic corona.
This is because of its large size, and its temperature is too high compared to the hottest temperatures \citep[1.5--2 keV;][]{sun07} observed in galactic corona associated with the most-massive elliptical galaxies.
The most similar dense cool-core system was found in A2107 by \citet{sun07}, with $r \sim 18$ kpc and $kT \sim 2.7$ keV, who suggested this hot and large component \citep[see also][]{fuj06} was most likely associated with the cluster cool-core.

Overall, the limited statistics of the current \Chandra\ exposure make it difficult to determine if there is significant X-ray substructure that may be associated with the finer details seen in the radio images. 
Near the center of the cluster, we found no radio emission coincident with the weak $\sim 2\sigma$ X-ray decrements at $\sim$8.5 kpc from the AGN found in the \Chandra\ images \citep{som18} that could support their interpretation as cavities.
Instead, the radio lobes in \PKS\ appear to be located just outside the X-ray cool-core (\S~\ref{radioXray}), a feature commonly found in cluster coronae with lower-power FR-I radio sources \citep[e.g.,][]{kra03,har05,sun05}. 
This anti-correlation of the radio and X-ray emission in the  FR-I/coronae systems could indicate the \textit{inner-most} kpc-scale jets in these systems do not significantly heat the coronae at smaller radii, but rather deposit the bulk of their energy at larger scales \citep{sun07}.

The 0.5--2\,keV luminosity of the $r=22$ kpc cool-core in \PKS, $L_{\rm X} = (1.2 \pm 0.1) \times 10^{43}$ erg s$^{-1}$ and its radio luminosity of $1.0 \times 10^{26}$ W Hz$^{-1}$ from the total NVSS 1.4 GHz flux density, makes it one of the most radio-luminous of the cool-core class \citep[][Figure~1, therein]{sun09}. 
At the same time, the X-ray size is smaller than typical cool-cores in clusters found in more nearby systems, and the cool-core is displaced $\sim$$2.5\arcsec$ ($\sim$9\,kpc) west of the AGN at the center of the cluster \citep{som18}.
The offset suggests gas motions (sloshing) inside the cool-core that can be induced by a gravitational perturbation of the cluster caused by a minor merger \citep[e.g.,][]{mar07}. 
The central AGN outburst may also contribute to sloshing by supplying kinetic energy to the surrounding cool gas \citep{qui01,mar01,hla11}. 
In this case, a pre-existing asymmetry in the X-ray gas distribution is likely needed to explain the different orientation of the radio outburst and the direction of the cool-core offset. 
Detailed simulations of AGN outbursts in asymmetric cool cores would be required to explore this possibility.

The AGN outburst could have heated the \textit{outer-most} layers of the cool core, leaving a compact, cool-core remnant \citep{sun04} as we observe. 
Sloshing may have further contributed to heating up the gas by facilitating heat inflow to the core \citep{zuh10}.
This provides a consistent picture with the similar angular scales of the observed small, offset, compact X-ray cool-core and the $\sim$20 kpc scale of the large-scale FR-II radio lobes revealed in the VLA maps.

\acknowledgments

We thank the anonymous referee for providing detailed comments on the manuscript.
We thank Lorant Sjouwerman for maintaining the NRAO NVAS database.
Portions of this work by C.C.C.\, at the Naval Research Laboratory were supported by the Chief of Naval Research. 
Basic research in radio astronomy (S.G., T.E.C.) at the Naval Research Laboratory is supported by 6.1 Base funding.

\vspace{5mm}
\facilities{Chandra(ACIS), VLA, VLBA}

\software{AIPS \citep{bri94}, CIAO \citep{fru06}, DIFMAP \citep{she94}, SAOImage DS9 \citep{joy03}, XSPEC \citep{dor01,dor03}
          }

\appendix

\section{VLBA Proper Motions}

\begin{figure}[t]
\begin{center}
\includegraphics[scale=0.55,angle=0]{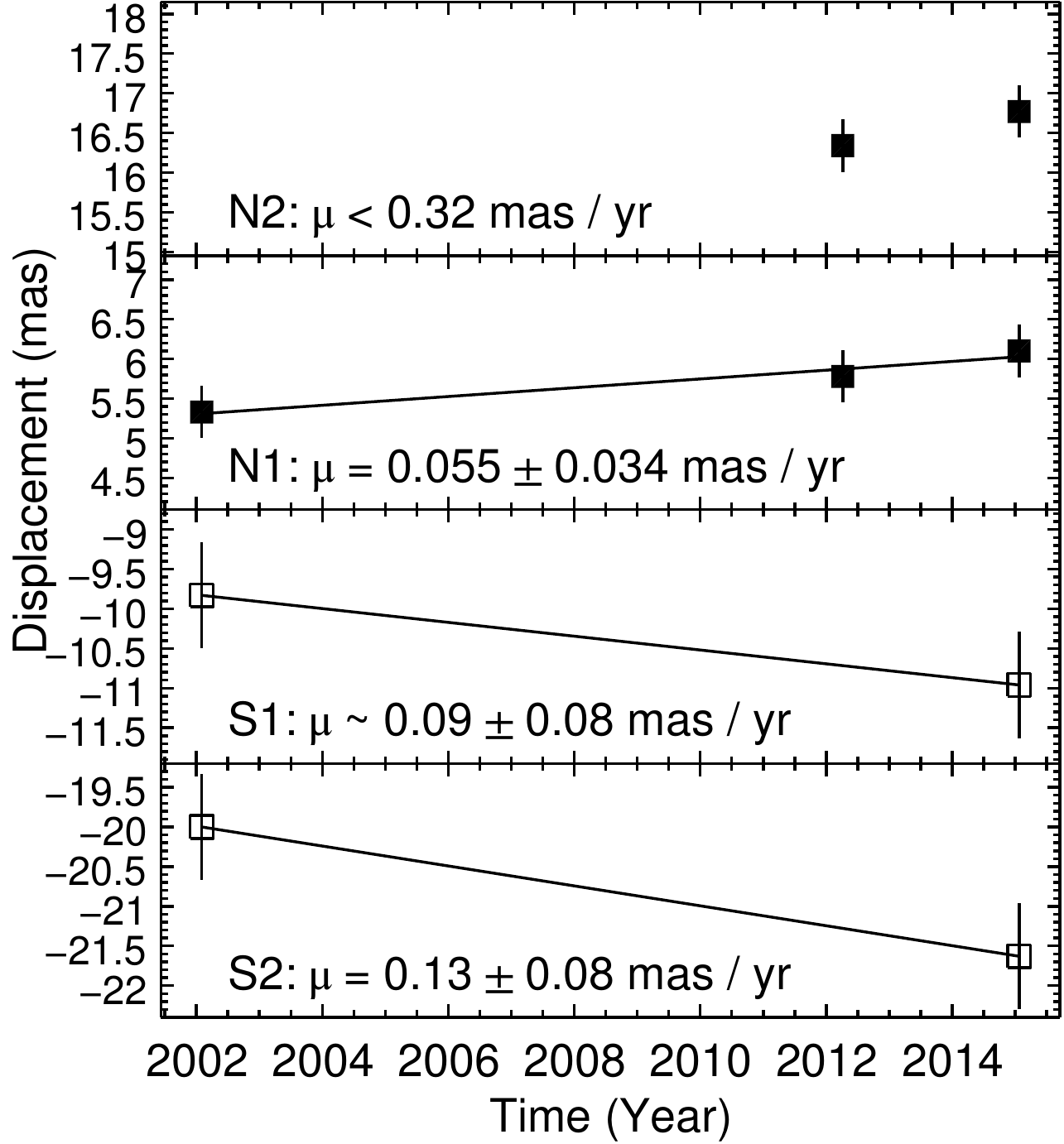}
\caption{Radial displacements of four distinct features in the VLBA jets relative to the core over time. The 8.4/8.7 GHz data are filled squares and the 2.3 GHz data are the open squares. The derived proper motions are indicated.}
\label{motions}
\end{center}
\end{figure}

The long, 13-year time spanned by the VLBA observations allow us to measure or usefully constrain the proper motions for common features in the southern jet seen in the 2.3 GHz images (S1 and S2) and in the northern jet seen in the 8.4/8.7 GHz images (N1 and N2). Their measured displacements relative to the core (Table~\ref{vlbaknots}) are plotted versus time in Figure~\ref{motions}. For the purpose of this analysis, uncertainties in the knot positions were assumed to be 1/6th of the common beam-widths \citep[see e.g.,][and references therein]{lis09} in the maps presented in Figure~\ref{vlbamontage}. At the source redshift, 0.07 mas yr$^{-1}$ = 1$c$.

The feature N1 is identifiable in all three 8.4/8.7 GHz datasets, with a clear proper motion detected. It is moving radially outward from the core at an average $PA = -24\deg$ with $\mu = 0.055 \pm 0.034$ mas yr$^{-1}$ that implies an apparent motion, $\beta_{\rm app} = v/c = 0.8 \pm 0.5$ for N1.

The remaining features are observed only in two epochs. The southern jet is detected only in the 2.3 GHz images with larger beam size than the ones at 8.7 GHz, so require larger proper motions for significant detections. Indeed, the largest proper motion is observed in the outer knot S2, moving along $PA = 154\deg$ at a rate, $\mu = 0.13\pm 0.08$ mas yr$^{-1}$ that implies apparent superluminal motion, $\beta_{\rm app} =1.9 \pm 1.1$.

The inner knot S1 has apparent substructure in the 4.5 mas-resolution 2.3 GHz images and the resultant fitted Gaussian component is elongated along the jet direction. There is a fitted component in the 2 mas-resolution 8.4 GHz image in the 2012 data whose position would be consistent with S1 seen at 2.3 GHz, but the additional elongated sub-structure in the higher-resolution image complicates the component identification and centroid measurement. With the two 2.3 GHz measurements, the proper motion is formally, $\mu \sim 0.09 \pm 0.08$ mas yr$^{-1}$ ($PA \sim 155\deg$), and would imply a superluminal motion as well, $\beta_{\rm app} \sim 1.3$.

Finally, the fainter, outer feature in the north jet, N2, is seen only in the latter two epochs. Its measurements are consistent with zero displacement ($\sim 0.4 \pm 0.5$ mas) along $PA \sim -22\deg$ that implies an upper limit of $\mu < 0.32$ mas yr$^{-1}$ ($1\sigma$).

\section{Historical Integrated Radio Spectrum}

The integrated radio measurements of \PKS\ are compiled from published data (Table~\ref{table-appendix}) predominantly compiled from NED\footnote{The NASA/IPAC Extragalactic Database (NED) is operated by the Jet Propulsion Laboratory, California Institute of Technology, under contract with the National Aeronautics and Space Administration.}. Additional simultaneous measurements in January 1997 from the VLA Calibrator Manual \citep{per03} are listed separately. The data are plotted in Figure~\ref{radiosed}.

A single-band measurement at 151 MHz of $1.579 \pm 0.065$ Jy was given for GLEAM J135605--342110 by \citet{hur17}, consistent with the 150 MHz GMRT measurement from the TGSSADR catalog \citep{int17}. The GLEAM data were however, obtained over a wide band, 72--231 MHz, and we measured the flux densities in four separate band images (72--103, 103--134, 139--170, and 170--231 MHz), obtained by Gaussian fits using {\tt JMFIT} in {\tt AIPS}. The measurements are presented at their corresponding central frequencies, with errors computed by adding a $5\%$ flux calibration uncertainty and the local image rms in quadrature. The individual band flux densities are mostly in line with other low-frequency measurements, except that the 87.5 MHz point is somewhat high, but the low-frequency spectral index seems overall consistent with $\alpha\sim0.5$. Taking the most-modern lowest-frequency measurements from the VLSSr 74 MHz (75\arcsec\ beam), TGSS 150 MHz (25\arcsec), and our three reliable GLEAM measurements at 118.5 ($\sim$3.5$'$), 154.5 ($\sim$2.6$'$), and 200.5 MHz ($\sim$2.2$'$), the best fit spectral index is, $\alpha = 0.55 \pm 0.12$. 

Note there is a known field radio source, NVSS J135553--341842, to the NW of \PKS, but we found no other radio detections of this faint  source ($8.3 \pm 0.5$ mJy at 1.4 GHz) in the other catalogs. The $3.6'$ offset between the sources is larger than the beam-sizes in the lower-frequency maps, and the NVSS source is likely too faint to have contributed to the integrated flux measurements of \PKS.

\begin{table*}
  \begin{center}
\caption{Historical integrated radio flux density measurements of \PKS.}
\begin{tabular}{lcc}
\hline
\hline
Frequency (GHz) & $S_{\nu}$ (mJy) & Catalog Name* \\
\hline
20 & 431 $\pm$ 21 & AT20G J135605--342111  \\ 
8.6 & 724 $\pm$ 36 & AT20G J135605--342111  \\ 
8.4 & 708 $\pm$ 70.8 & CRATES J135604--342041*  \\ 
5 & 670 $\pm$ 67 & PKS J1356--3421*  \\ 
4.85 & 731 $\pm$ 52 & PMN J1356--3420  \\ 
4.8 & 690 $\pm$ 34 & AT20G J135605--342111  \\ 
2.7 & 604 $\pm$ 64 & PKS J1356--3421*  \\ 
2.3 & 560.9 $\pm$ 13.7 & SPASS J135603--342101  \\ 
2.29 & 640 $\pm$ 64 & \citet{jau82}* \\
1.4 & 691.8 $\pm$ 20.8 & NVSS J135605--342110  \\ 
1.4 & 692 $\pm$ 69.2 & AT20G-harc J135605--342111*  \\ 
0.843 & 843.2 $\pm$ 25.4 & SUMSS J135605--342109  \\ 
0.365 & 1341 $\pm$ 104 & TXS 1353--341  \\ 
0.408 & 940 $\pm$ 94 & PKS J1356--3421*  \\ 
0.408 & 900 $\pm$ 70 & MRC 1353--341  \\ 
0.2005 & 1292 $\pm$ 68 & GLEAM J135605--342110 / This work \\ 
0.1545 & 1513 $\pm$ 84 & GLEAM J135605--342110 / This work \\ 
0.15 & 1516.5 $\pm$ 151.9 & TGSSADR J135605.3--342109  \\ 
0.1185 & 1760 $\pm$ 122 & GLEAM J135605--342110 / This work \\ 
0.0875 & 2797 $\pm$ 168 & GLEAM J135605--342110 / This work \\ 
0.074 & 2160 $\pm$ 300 & VLSSr J135604.8--342114  \\ 
\hline
43 & 240 $\pm$ 24 & VLA Calibrator Manual*   \\ 
15 & 600 $\pm$ 60 & VLA Calibrator Manual*  \\ 
8.1 & 690 $\pm$ 69 & VLA Calibrator Manual*   \\ 
5 & 780 $\pm$ 78 & VLA Calibrator Manual*   \\ 
1.5 & 680 $\pm$ 68 & VLA Calibrator Manual*   \\ 
\hline
\end{tabular}
\end{center}
\smallskip
*Catalogs / Names indicated with asterisks did not quote uncertainties, and $10\%$ was assumed here.\\
References: 
AT20G \citep{mur10},
AT20G-harc \citep{chh13}, 
CRATES \citep{hea07},
GLEAM \citep{hur17}, 
\citet{jau82},
NVSS \citep{con98},
PKS \citep{wri90}, 
PMN \citep{wri94}, 
SPASS \citep{mey17}, 
SUMSS \citep{mau03}, 
TGSSADR with GMRT \citep{int17}, 
TXS \citep{dou96},
and
VLSSr \citep{lan14} from \url{https://www.cv.nrao.edu/vlss/VLSSlist.shtml}.
\\
\label{table-appendix}
\end{table*}

\end{document}